\documentclass[
aps,
fleqn,
showkeys,
twocolumn
]{revtex4} 
\usepackage{graphicx}
\usepackage{amssymb}
\usepackage{amsmath}
\usepackage{xcolor}
\usepackage{bm}
\usepackage[T1]{fontenc}
\usepackage[normalem]{ulem} 

\begin{document}
\title[]{Impact of Negative Weights on Performance in Reservoir Computing}

\author{Yeonsu \surname{Jeong}}
\thanks{These authors contributed equally}
\affiliation{Department of Applied Physics, Hanyang University, Ansan 15588, Korea}
\author{Seonho \surname{Nam}}
\thanks{These authors contributed equally}
\affiliation{Department of Applied Physics, Hanyang University, Ansan 15588, Korea}
\affiliation{R\&D Center, Green Star Incorporated, Anyang 14056, Korea}
\author{Dahan \surname{Choi}}
\thanks{These authors contributed equally}
\affiliation{Department of Applied Physics, Hanyang University, Ansan 15588, Korea}
\affiliation{R\&D Center, Green Star Incorporated, Anyang 14056, Korea}
\author{Mi Jin \surname{Lee}}
\email{mijinlee@pusan.ac.kr}
\affiliation{Department of Physics, Pusan National University, Busan 46241, Korea}
\author{Seung-Woo \surname{Son}}
\email{sonswoo@hanyang.ac.kr}
\affiliation{Department of Applied Physics, Hanyang University, Ansan 15588, Korea}

\newcommand{\noteYeonsu}[1]{\textbf{\textcolor{blue}{{\footnotesize{Yeonsu: }}#1}}}

\begin{abstract}
Reservoir computing provides an efficient framework for time-series prediction because only the readout layer is trained while the recurrent reservoir remains fixed. In this study, we examine how the negative sign ratio of a signed Erd\H{o}s--R\'enyi reservoir affects the internal dynamics and forecasting performance of an echo state network on the Mackey--Glass benchmark. To characterize the reservoir responses, we use two Shannon-entropy-based observables, namely node-state entropy and node-frequency entropy, as empirical proxies for the diversity of reservoir states and temporal activation patterns. We find that intermediate negative sign ratios are associated with richer internal dynamics and substantially lower prediction error than the sign-homogeneous extremes for the task considered here. These results suggest that link-sign composition can be a useful structural parameter in reservoir design for chaotic forecasting. The present findings should be interpreted within the scope of the current benchmark and reservoir setting, and broader validation will be required to establish their generality across other tasks, topologies, and controlled spectral conditions.

\end{abstract}


\maketitle

\section{INTRODUCTION}

Artificial neural networks (ANNs) have become increasingly ubiquitous across diverse domains of scientific and technological applications, fundamentally transforming fields ranging from autonomous vehicle navigation to computational creativity in image synthesis algorithms~\cite{diraco2019ai,cai2021robot,angelov2021explainable,tanaka2019recent}. Their computational capabilities extend to natural language processing and environmental modeling, where they contribute significantly to the accuracy and reliability of prediction systems~\cite{wingfield2022similarities,kalhor2020introduction,huang2021predicting,degrave2022magnetic}. Recent advances in representation learning have further enabled efficient retrieval and similarity search in large-scale document corpora, such as patents, by embedding textual data into latent spaces using pretrained language models and contrastive learning techniques~\cite{Jo2023contrastive}. The exponential growth in computational resources and data availability has facilitated the evolution of increasingly sophisticated ANN architectures~\cite{kalhor2020introduction,lorencin2020using,lyu2022prediction}. This technological advancement has enabled the development of deep learning models capable of extracting intricate patterns and generating predictions with unprecedented accuracy, including representation learning frameworks that embed complex user characteristics into latent spaces using metric learning approaches such as triplet loss~\cite{Koo2023triplet}. Nevertheless, the substantial computational demands and extensive training time required by state-of-the-art ANN architectures remain significant challenges, particularly in resource-constrained computational environments.

\begin{figure*}
    \centering
    \includegraphics[width=0.95\linewidth]{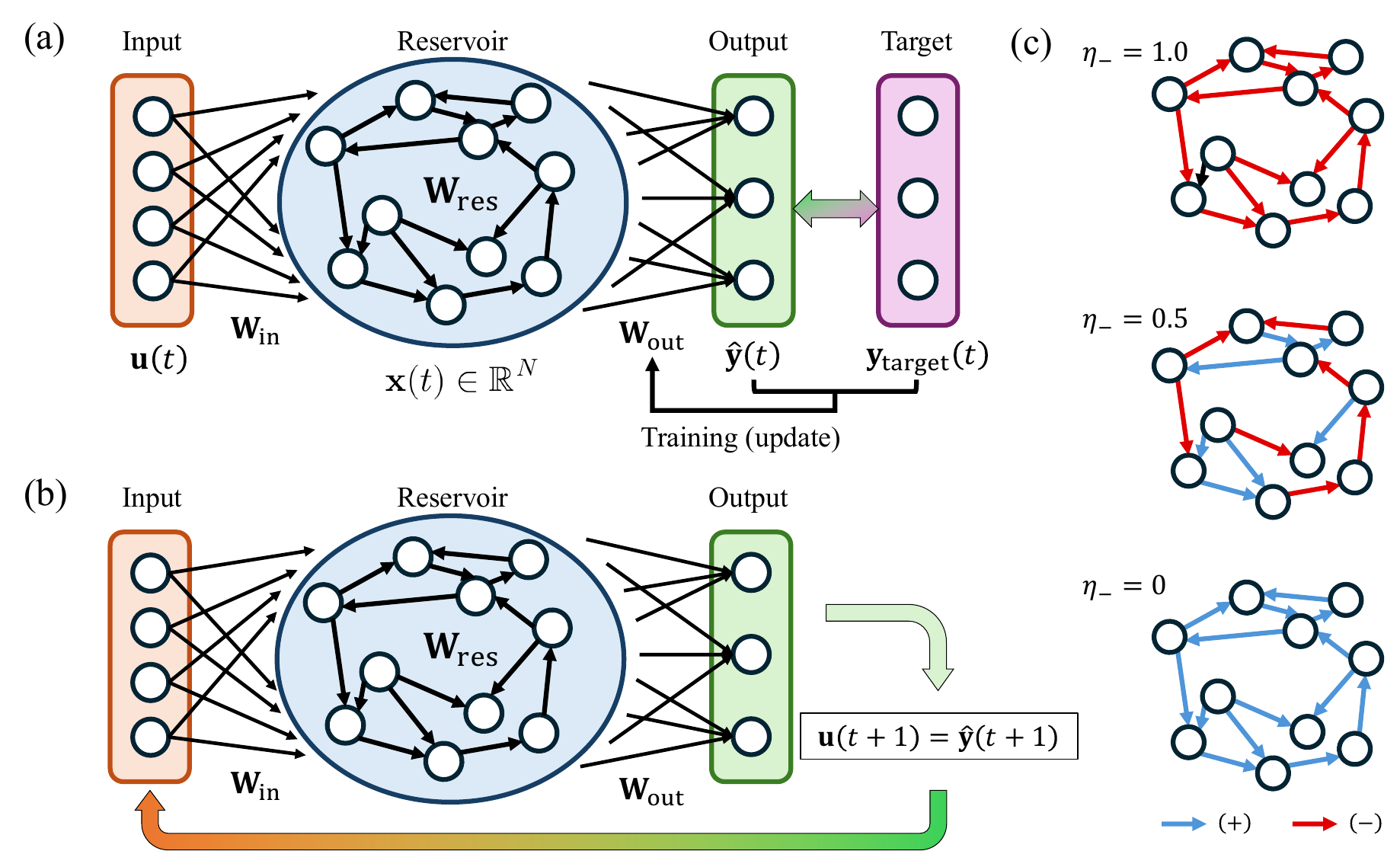}
    \caption{Schematic architecture of the RC model. (a) Training mode, in which the input time series is fed into the reservoir and the readout weights $\mathbf{W}_{\rm out}$ are optimized by comparing the output with the ground truth. (b) Prediction mode, in which the model operates autonomously by feeding the previous output back as the next input for iterative multi-step forecasting. (c) Construction of signed reservoir networks with controlled negative sign ratios $\eta_{-}$ by systematically reassigning the signs of links in an initial Erd\H{o}s--R\'enyi (ER) reservoir.}
    \label{fig:rc_architecture}
\end{figure*}

Within this context, reservoir computing (RC) has emerged as a compelling computational paradigm offering distinct advantages. Characterized by reduced computational overhead and efficient training protocols, RC presents a robust alternative to conventional ANN architectures~\cite{lukovsevivcius2009reservoir,jaeger2001echo,kawai2019small,tanaka2019recent}. The defining feature of RC, a fixed reservoir network structure maintained throughout the learning phase, enables a streamlined training procedure that circumvents the need for extensive iterative weight optimization characteristic of deep learning approaches~\cite{zhang2021power,lukovsevivcius2009reservoir}. Furthermore, the architectural stability inherent to RC proves particularly advantageous for physical neural network implementations, where hardware constraints impose limitations on system flexibility and adaptability. This fixed architecture substantially simplifies the physical realization of ANNs, establishing RC as a promising methodology for developing computationally efficient artificial intelligence systems deployable across diverse applications, from edge computing devices to large-scale industrial systems~\cite{diraco2019ai,kalhor2020introduction,huang2021predicting,zhang2021power,wingfield2022similarities,degrave2022magnetic}.

Among the major RC paradigms, echo state networks (ESNs) are particularly attractive for chaotic time-series prediction because only the readout layer is trained while the recurrent reservoir remains fixed~\cite{jaeger2001echo,lukovsevivcius2009reservoir}. This separation between nonlinear dynamical transformation and lightweight readout learning makes RC highly effective for tasks in which temporal memory and nonlinear mixing are both essential~\cite{gauthier2021next,vlachas2020backpropagation,pathak2018model}. Figure~\ref{fig:rc_architecture} summarizes the workflow adopted in this study: the target signal is injected during training, multi-step forecasting is performed by autonomous feedback during prediction, and the reservoir itself is systematically converted into a signed network by controlling the fraction of negative links.
Although random reservoirs are often treated as generic dynamical substrates, their detailed sign structure can strongly influence the resulting internal dynamics. In signed directed networks, positive and negative connections encode qualitatively different modes of interaction, and changing their ratio alters the balance between amplification and suppression in the network response~\cite{zaslavsky1982signed,kunegis2010spectral,guan2023edge,du2023abc}. In the context of RC, such structural changes are expected to modify the reservoir's eigenvalue distribution, state diversity, temporal complexity, and ultimately its predictive utility~\cite{brunetti2022unbalanced,verstraeten2009quantification}. However, the role of the negative sign ratio in reservoir design has not been systematically clarified.

In this work, we investigate how the negative sign ratio of an ER reservoir affects the dynamics and predictive performance of an ESN on the Mackey--Glass benchmark. We combine dynamical inspection of the reservoir states with two complementary Shannon-entropy measures, namely node-state entropy and node-frequency entropy, to quantify the information capacity induced by different sign compositions. We then relate these entropy measures to forecasting accuracy. Our central finding is that an intermediate negative sign ratio generates richer internal dynamics and yields substantially lower prediction error than either sign-homogeneous extreme. This result provides a simple but effective structural principle for improving reservoir design. 

\begin{figure*}[t]
    \centering
    \includegraphics[width=1\linewidth]{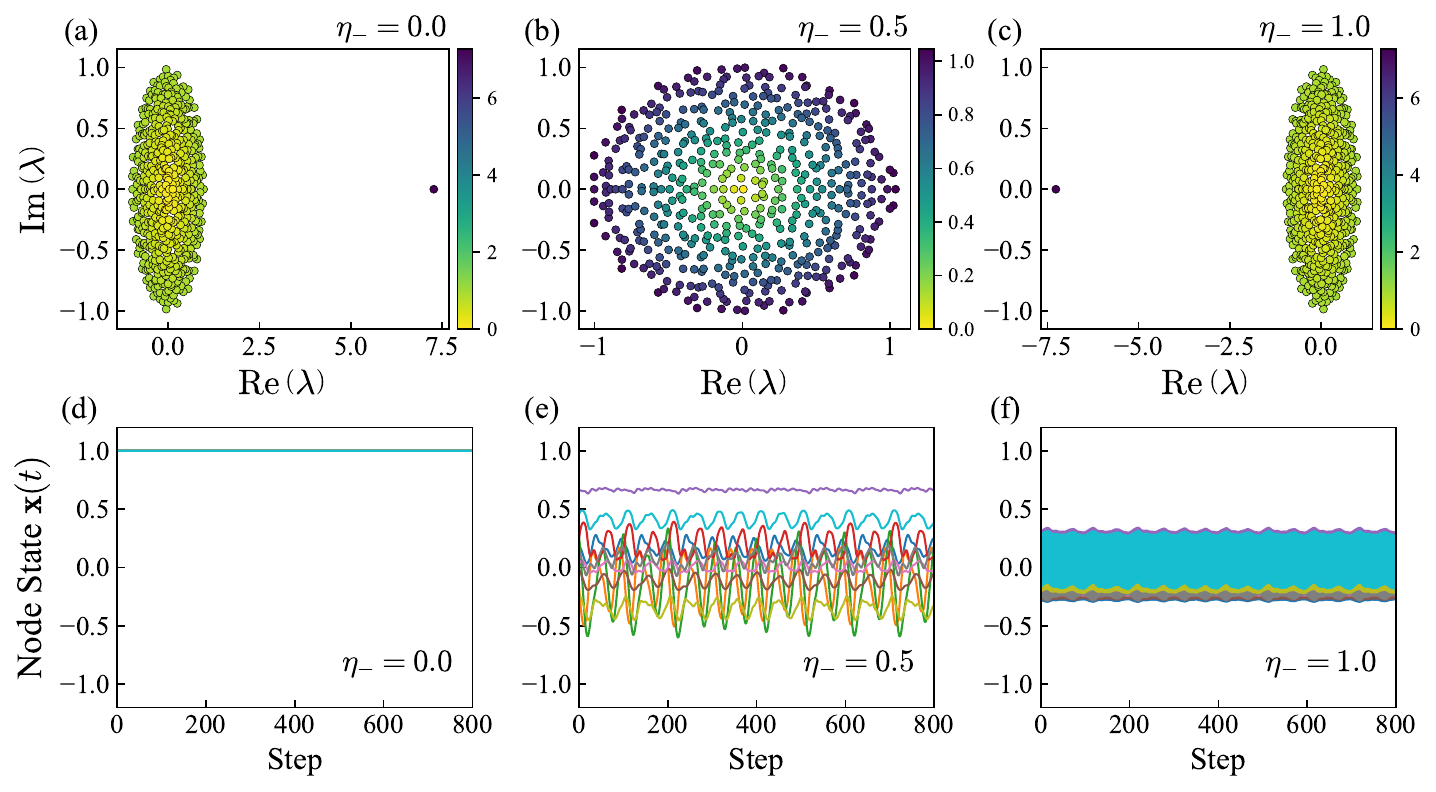}
    \caption{Effect of the negative sign ratio on the internal dynamics of the reservoir. (a)-(c) Eigenvalue distributions of the reservoir weight matrix in the complex plane for representative negative sign ratios (0.0, 0.5, and 1.0), showing how the spectral structure changes with link-sign composition. (d)-(f) Representative node-state trajectories for the corresponding reservoirs, illustrating that intermediate sign ratios generate richer and more heterogeneous temporal responses than the extreme cases.}
    \label{fig:spectrum_states}
\end{figure*}

\section{EXPERIMENTAL METHODOLOGY}
\emph{Echo-state network} - Reservoir computing is commonly implemented using two major paradigms: ESNs and liquid state machines. In this work, we focus on the ESN framework because it enables efficient training with a fixed recurrent reservoir, while only the readout parameters are learned~\cite{jaeger2001echo,lukovsevivcius2009reservoir}. The ESN architecture consists of three functionally distinct layers: an input layer, a recurrent reservoir, and an output layer. As illustrated in Fig.~\ref{fig:rc_architecture}(a), the input signal $\mathbf{u}(t)\in\mathbb{R}^{N_u}$ is injected into the reservoir, generating the internal state $\mathbf{x}(t)\in\mathbb{R}^{N}$. In the present work, the reservoir dynamics are governed by

\begin{equation}
\begin{split}
    \mathbf{x}(t)
    &=(1-\alpha)\mathbf{x}(t-1) ~ \\[3pt]
    &~+\alpha \tanh\!\left(
    \mathbf{W}_{\rm in}\mathbf{u}(t)
    +\mathbf{W}\mathbf{x}(t-1)
    +\mathbf{b}_{\rm res}
    \right),
\end{split}
\end{equation}
where $\alpha$ is the leaking rate, $N$ is the number of reservoir nodes, $\mathbf{W}_{\rm in}\in\mathbb{R}^{N\times N_u}$ is the input weight matrix, $\mathbf{W}\in\mathbb{R}^{N\times N}$ is the recurrent reservoir weight matrix, and $\mathbf{b}_{\rm res}\in\mathbb{R}^{N}$ is the reservoir bias vector.

The one-step-ahead prediction is obtained from an augmented feature vector consisting of the current input and reservoir state,

\begin{equation}
    \hat{\mathbf{y}}(t+1)
    =
    \widetilde{\mathbf{W}}_{\rm out}
    \widetilde{\mathbf{x}}(t)~,
\end{equation}
where the augmented readout matrix and feature vector are defined as

\begin{equation}
\begin{split}
\widetilde{\mathbf{W}}_{\rm out}
&=
\begin{bmatrix}
\mathbf{b}_{\rm out} &
\mathbf{W}_{u} &
\mathbf{W}_{\rm out}
\end{bmatrix}
\in
\mathbb{R}^{N_y\times(1+N_u+N)}, \\[8pt]
\widetilde{\mathbf{x}}(t)
&=
\begin{bmatrix}
1\\
\mathbf{u}(t)\\
\mathbf{x}(t)
\end{bmatrix}
\in
\mathbb{R}^{1+N_u+N}.
\end{split}
\end{equation}

Here, $\mathbf{b}_{\rm out}\in\mathbb{R}^{N_y}$ is the readout bias, $\mathbf{W}_{u}\in\mathbb{R}^{N_y\times N_u}$ represents the direct contribution of the input to the output, and $\mathbf{W}_{\rm out}\in\mathbb{R}^{N_y\times N}$ denotes the readout weights associated with the reservoir state. In this study, both the input and output are scalar variables ($N_u=N_y=1$), since the Mackey--Glass benchmark is represented by a univariate time series.


The training and prediction modes differ in how the reservoir is driven. During training, the target time series is provided, and the readout parameters are determined from the augmented features constructed from the input and reservoir states, as shown in Fig.~\ref{fig:rc_architecture}(a). Unlike conventional recurrent neural-network training based on backpropagation, in which network parameters are iteratively updated, the readout parameters of the ESN are obtained directly through ridge regression without iterative gradient-based optimization. During prediction, the model operates autonomously: the predicted output at each time step is fed back as the input for the next reservoir update, $\mathbf{u}(t+1)=\hat{\mathbf{y}}(t+1)$, as shown in Fig.~\ref{fig:rc_architecture}(b). This closed-loop configuration enables iterative multi-step forecasting and provides a direct test of whether the reservoir has captured the underlying temporal dynamics of the chaotic signal.

\begin{figure*}[t]
    \centering
    \includegraphics[width=\linewidth]{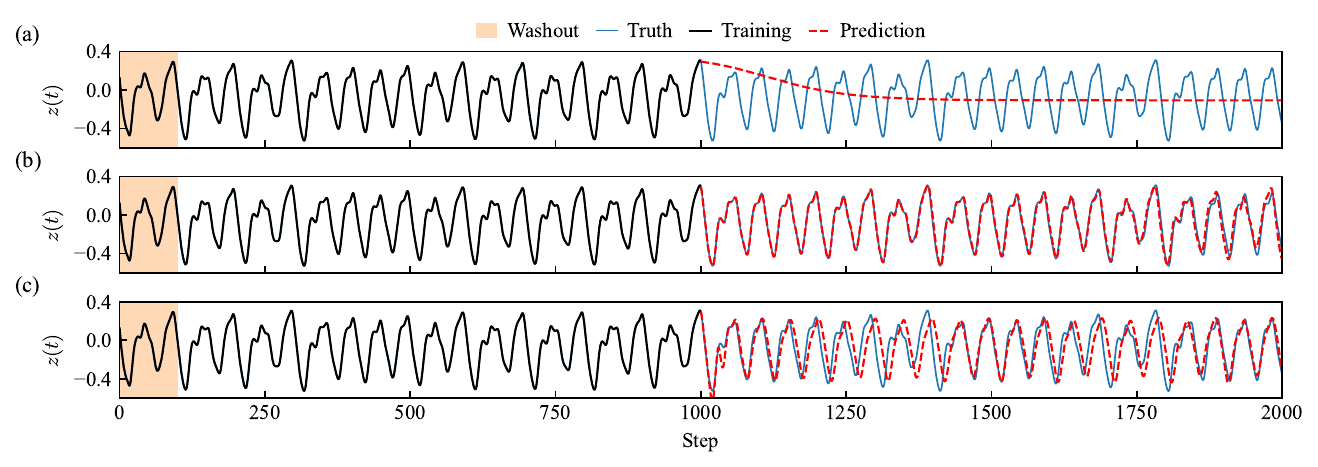}
    \caption{
    Temporal evolution of the Mackey--Glass time series and the corresponding ESN predictions for different negative sign ratios: (a) \(\eta_{-}=0\), (b) \(\eta_{-}=0.5\), and (c) \(\eta_{-}=1\). The shaded region indicates the washout period, while the black, blue, and red dashed lines denote the training data, ground truth, and autonomous prediction, respectively.
    }
    \label{fig:mackeyglass}
\end{figure*}

For the readout layer, we employ ridge regression, which provides a simple and numerically stable approach for fitting the linear readout while mitigating overfitting~\cite{kawai2019small,pathak2018model,wyffels2008stable}. During training, the input sequence is sequentially fed into the reservoir to generate the corresponding reservoir states. After discarding the initial washout period, the retained input and reservoir state at each time step are combined into the augmented feature vector $\widetilde{\mathbf{x}}(t)$. These feature vectors are collected column-wise into the augmented feature matrix
\begin{equation}
\widetilde{\mathbf{X}}
=
\begin{bmatrix}
\widetilde{\mathbf{x}}(t_1) &
\widetilde{\mathbf{x}}(t_2) &
\cdots &
\widetilde{\mathbf{x}}(t_{T_{\rm tr}})
\end{bmatrix}
\in
\mathbb{R}^{(1+N_u+N)\times T_{\rm tr}},
\end{equation}
where $T_{\rm tr}$ denotes the number of training samples retained after the washout period. Since the readout is trained for one-step-ahead prediction, the corresponding target values are collected as
\begin{equation}
\mathbf{Y}_{\rm target} =
\left[
\mathbf{y}_{\rm target}(t_k+1)
\right]_{k=1}^{T_{\rm tr}}
\in
\mathbb{R}^{N_y\times T_{\rm tr}},
\end{equation}
where \(t_k\) denotes the \(k\)-th retained time index. The augmented readout matrix is then obtained by ridge regression as
\begin{equation}
\widetilde{\mathbf{W}}_{\rm out}
=
\mathbf{Y}_{\rm target}
\widetilde{\mathbf{X}}^{\mathrm T}
\left(
\widetilde{\mathbf{X}}\widetilde{\mathbf{X}}^{\mathrm T}
+
\beta\mathbf{I}
\right)^{-1},
\label{eq:readout_update}
\end{equation}
where $\beta$ is the regularization parameter and
$\mathbf{I}\in\mathbb{R}^{(1+N_u+N)\times(1+N_u+N)}$
is the identity matrix. This readout-only training strategy leaves the reservoir parameters fixed while allowing the output mapping to be determined efficiently by linear regression.

\emph{Signed reservoir} - The reservoir network in our RC model is constructed from an ER random graph with $N=500$ nodes and connection probability $p=0.1$~\cite{wyffels2010comparative,pathak2018model,lukovsevivcius2009reservoir,kawai2019small}. While the classical ER model specifies only the existence of links, the reservoir implementation additionally includes directionality and link weights~\cite{erdHos1960evolution,son2009dynamics,zeng2011enhancing}. In conventional reservoir construction, the reservoir weight matrix $\mathbf{W}$ is commonly rescaled by its spectral radius, defined as the largest absolute eigenvalue. In the present study, however, $\mathbf{W}$ is normalized using the magnitude of the second-largest eigenvalue. Normalization by the largest eigenvalue, i.e., the spectral radius, produces slowly varying reservoir states and a correspondingly long transient relaxation in the present setting. The use of the second-largest eigenvalue suppresses this slow transient behavior. The signs of a prescribed fraction of links are subsequently reversed, with no additional normalization after the sign reassignment. The spectrum is therefore allowed to vary with the sign composition.


We define the negative sign ratio as
\begin{equation}
    \eta_{-}=\frac{N_{-}}{E},
\end{equation}
where $N_-$ is the number of negative-sign links and $E$ is the total number of links in the reservoir network. Figure~\ref{fig:rc_architecture}(c) schematically illustrates this construction procedure. The sign ratio is varied from 0 to 1 in increments of 0.05, with $\eta_{-}=0$ corresponding to an all-positive reservoir and $\eta_{-}=1$ corresponding to an all-negative reservoir. Because the sign structure directly changes the spectrum of the recurrent weight matrix, it also changes the balance between stability, diversity, and responsiveness in the reservoir dynamics~\cite{brunetti2022unbalanced,verstraeten2009quantification}. The aim of the present work is to quantify how these changes propagate to entropy and prediction performance.

\emph{Chaotic benchmark} - The Mackey--Glass system, introduced in 1977 by Mackey and Glass, is a nonlinear time-delay dynamical system originally formulated to model hematopoietic regulation in biological systems~\cite{mackey1977oscillation}. In the present study, 
the resulting scalar time series is used as both the input to the reservoir during training and the target for prediction. The governing equation is
\begin{equation}
\frac{dz}{dt}=\frac{0.2z(t-\tau)}{1+z(t-\tau)^{10}}-0.1z(t)~,
\end{equation}
where the delay parameter is fixed at $\tau=17$, for which the system exhibits chaotic behavior~\cite{junges2012intricate}, and the resulting time series is illustrated in blue lines in Fig.~\ref{fig:mackeyglass}(a-c). The signal displays irregular oscillations with nonrepeating amplitudes and intervals, making it a demanding benchmark for recurrent predictors. This task is therefore well suited for testing whether a given reservoir design can simultaneously retain short-term memory and perform nonlinear transformation of past inputs.

\emph{Entropy measures and performance metric} - To quantify the effect of the signed reservoir structure on its internal dynamics, we analyze two complementary measures based on Shannon entropy~\cite{shannon1948mathematical}: node-state entropy and node-frequency entropy. 
{For a discrete probability distribution $\{p_i\}$ obtained by binning an observable, the Shannon entropy is defined as
\begin{equation}
    H=-\sum_i p_i \log p_i~.
    \label{eq:def_entropy}
\end{equation}
\label{eq:def_state_entropy}
The two entropy measures considered here differ in the observable used to construct the probability distribution and in their corresponding binning procedures. 
The node-state entropy characterizes the instantaneous diversity of reservoir states across nodes~\cite{mandal2021achieving}. At the observation time $T$, the state values $\{x_i(T)\}_{i=1}^{N}$ of all reservoir nodes are divided into $B_{\rm state}=40$ equally spaced bins over the interval $[-1,1]$. For the $j$-th bin, the probability is defined as $p_j^{\rm state}=n_j^{\rm state}/N$, where $n_j^{\rm state}$ is the number of nodes whose state falls within that bin. The node-state entropy is then given by
\begin{equation}
    H_{\rm state}
    =
    -\sum_{j=1}^{B_{\rm state}}
    p_j^{\rm state}\log p_j^{\rm state}~.
\end{equation}
A larger $H_{\rm state}$ therefore indicates that the reservoir nodes occupy a broader range of state values at the observation time.
}

{The node-frequency entropy characterizes the diversity of dominant temporal frequencies across reservoir nodes. For each node $i$, the power spectrum is obtained from its temporal state trajectory, and the peak frequency $f_i^{\rm peak}$ is identified as the frequency at which the spectral power is maximal. The absolute peak frequencies $\{|f_i^{\rm peak}|\}_{i=1}^{N}$ are divided into $B_{\rm freq}=20$ equally spaced bins spanning the minimum and maximum values of the corresponding peak-frequency distribution. For the $j$-th frequency bin, the probability is defined as $p_j^{\rm freq}=n_j^{\rm freq}/N$, where $n_j^{\rm freq}$ is the number of nodes whose peak frequency falls within that bin. The node-frequency entropy is then given by
\begin{equation}
    H_{\rm freq}
    =
    -\sum_{j=1}^{B_{\rm freq}}
    p_j^{\rm freq}\log p_j^{\rm freq}.
\end{equation}
Thus, $H_{\rm state}$ describes the instantaneous heterogeneity of node states, whereas $H_{\rm freq}$ describes the heterogeneity of dominant temporal frequencies across the reservoir.
}

The predictive performance of the ESN is evaluated by the mean squared error (MSE) between the target signal and the autonomous multi-step prediction. By comparing the MSE with the entropy measures across sign ratios, we assess whether a richer internal dynamical repertoire indeed translates into better forecasting of chaotic dynamics.

\section{RESULTS AND DISCUSSION}


\emph{Signed reservoirs reshape spectral and state dynamics} - We first examine how the sign composition modifies the intrinsic dynamics of the reservoir. Figure~\ref{fig:spectrum_states}(a) shows representative eigenvalue distributions of the reservoir weight matrix for $\eta_{-}=0.0$, $0.5$, and $1.0$. When the reservoir is dominated by one sign, the eigenvalue cloud is strongly elongated and accompanied by isolated real outliers, indicating biased dynamical modes. In contrast, the mixed-sign case $\eta_{-}=0.5$ produces a more balanced and dispersed spectrum in the complex plane. This more isotropic spectral structure implies that no single dynamical direction dominates the reservoir response.

The dynamical consequence of this spectral reorganization is evident in Fig.~\ref{fig:spectrum_states}(b). For $\eta_{-}=0.0$, many node trajectories occupy similar narrow bands, indicating limited heterogeneity across nodes. For $\eta_{-}=1.0$, several node states are driven toward nearly fixed or weakly varying values, suggesting partial saturation of the reservoir response. The intermediate sign ratio, however, produces node activations with clearly different amplitudes, phases, and temporal patterns. In other words, mixing positive and negative links promotes a broader set of internal dynamical modes. These observations show that changes in the link-sign composition are accompanied by substantial changes in both the spectral structure of the reservoir and its dynamical responses.


\begin{figure*}[th]
    \centering
    \includegraphics[width=0.95\linewidth]{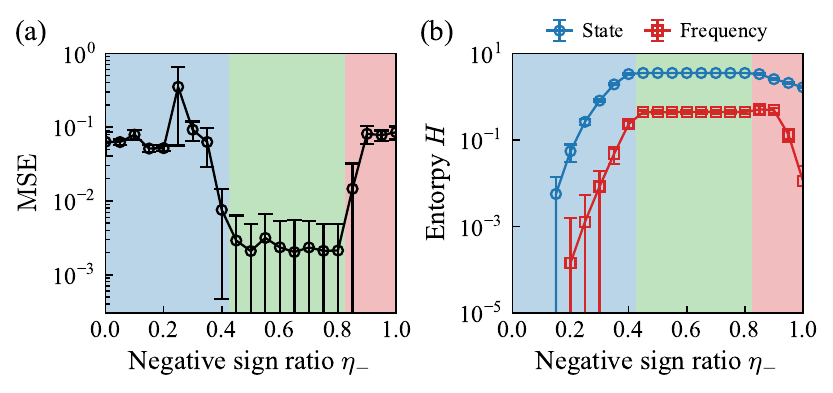}
    \caption{
    Prediction performance and entropy-based measures versus $\eta_{-}$. (a) Circles indicate the mean MSE, and error bars denote the standard deviation. (b) The colored symbols denote node-state entropy $H_{\rm state}$ (blue) and $H_{\rm freq}$ (red), and error bars denote the standard deviation.  Shaded regions are included as visual guides to distinguish different ranges of $\eta_{-}.$
    }
    \label{fig:perfom_and_entropy}
\end{figure*}

\emph{Prediction performance exhibits an intermediate-sign optimum} - 
To evaluate the forecasting performance of the ESN as a function of the negative sign ratio, we calculate the MSE, as shown in Fig.~\ref{fig:perfom_and_entropy}(a). For $\eta_{-} \lesssim 0.425$, the MSE remains at a comparable level with no pronounced dependence on $\eta_{-}$. In the intermediate range $(0.425 \lesssim \eta_{-} \lesssim 0.825)$, the MSE decreases sharply by several orders of magnitude, indicating substantially improved forecasting performance. For $\eta_{-} \gtrsim 0.825$, the MSE increases again and returns to values comparable to those observed at $\eta_{-} \lesssim 0.425$. Thus, the lowest prediction errors are obtained in the mixed-sign regime rather than at either sign-homogeneous limit. This nonmonotonic dependence indicates that the link-sign composition, together with the accompanying changes in reservoir dynamics, is strongly associated with forecasting performance in the present setting. The pronounced reduction in MSE over the intermediate range therefore motivates a closer examination of the reservoir dynamics associated with this low-error regime.

\emph{Entropy characterizes the low-error dynamical regime} - 
To characterize the dynamical properties of this low-error regime, Fig.~\ref{fig:perfom_and_entropy}(b) shows the node-state and node-frequency entropies as functions of $\eta_{-}$. The node-state entropy $H_{\rm state}$ remains low at small $\eta_{-}$ and increases sharply as negative links are introduced. For $\eta_{-} \gtrsim 0.425$, $H_{\rm state}$ remains high over a broad mixed-sign range, followed by a slight decrease for $\eta_{-} \gtrsim 0.825$. This behavior indicates that individual reservoir nodes explore a broader range of state values over time as the sign composition becomes mixed. The node-frequency entropy exhibits a stronger dependence on the negative sign ratio. It is relatively large over an intermediate range of $\eta_{-}$ but decreases markedly for $\eta_{-} \gtrsim 0.825$. Thus, the reservoir can maintain a relatively broad of visited state values even when the diversity of dominant frequencies across nodes becomes substantially reduced. The two entropy measures therefore characterize complementary aspects of the reservoir dynamics: the diversity of state values explored by individual nodes and the heterogeneity of their dominant temporal frequencies.

Comparison with the prediction results in Fig.~\ref{fig:perfom_and_entropy}(a) shows that the low-error region broadly coincides with high node-state entropy and appreciable node-frequency entropy. For $\eta_{-} \gtrsim 0.825$, however, the node-state entropy remains relatively high while the node-frequency entropy decreases and the prediction error increases. This observation indicates that broad exploration of node-state values alone is not sufficient to characterize the regime of accurate forecasting. Within the present setting, low prediction error is associated with the coexistence of broad state exploration and heterogeneous dominant temporal frequencies.

These results suggest that entropy-based measures can provide useful diagnostics of changes in reservoir dynamics induced by the link-sign composition. For the Mackey--Glass task considered here, the two entropy measures help characterize the dynamical regime associated with improved forecasting performance. They should, however, be regarded as empirical indicators of state and frequency diversity rather than as general measures of computational capacity.

\section{CONCLUSION}
In this study, we investigated how the negative sign ratio of a signed ER reservoir influences the internal dynamics and prediction performance of an echo state network on the Mackey--Glass benchmark. Within the present setting, reservoirs with intermediate negative sign ratios exhibited more heterogeneous node dynamics, larger entropy-based diversity measures, and lower prediction error than the sign-homogeneous extremes. These observations indicate that link-sign composition is closely associated with the dynamical and predictive behavior of the reservoir in the present setting.

At the same time, the present work should be interpreted as a proof of concept rather than as a general design law for RC. The analysis was performed on a single benchmark task with a specific reservoir construction, and the entropy-based measures used here are best understood as empirical proxies for internal state and dominant-frequency diversity rather than formal measures of computational capacity. Therefore, our results support the practical relevance of link-sign composition in the current setting, but they do not yet establish universal optimal sign ratios for reservoir design.

Several directions for future studies follow naturally from this work. First, it will be important to test whether the intermediate-sign optimum persists when the spectral radius is held fixed across different sign ratios, thereby partially separating the effect of sign composition from variations in spectral radius. Second, the robustness of the present result should be further examined across different reservoir sizes, sparsities, and input scalings. Third, the same framework should be evaluated on additional benchmark tasks, including other chaotic systems and real-world temporal datasets, to determine how task-dependent the optimal sign regime is. Fourth, entropy-based indicators should be compared with more formal measures such as memory capacity and related computational-capacity metrics~\cite{jaeger2007echo,dambre2012information}. Finally, extending the present framework to alternative reservoir topologies and hardware-oriented physical implementations may clarify when signed interactions provide a robust practical advantage.

Overall, the present study suggests that the negative sign ratio is a promising structural parameter for shaping reservoir dynamics in chaotic prediction, and motivates broader investigations of signed reservoirs across reservoir-computing tasks and architectures.

\begin{acknowledgments}
This work was supported by the National Research Foundation (NRF) of Korea through the Grant Nos. RS-2024-00341317 (M.J.L.) and RS-2026-25488703 (S.-W.S.). 
We also acknowledge the hospitality at APCTP, where part of this work was done. We thank APCTP, Pohang, Korea, for their hospitality during the Topical Research Program [APCTP-2025-T04], from which this work greatly benefited. 
\end{acknowledgments}


\end{document}